# Complete Strain Mapping of Nanosheets of Tantalum Disulfide

Yue Cao,*,†,‡ Tadesse Assefa,† Soham Banerjee,†,¶ Andrew Wieteska,§ Dennis Zi-Ren Wang,¶ Abhay Pasupathy,§ Xiao Tong,& Yu Liu,// Wenjian Lu,// Yu-Ping Sun,//,⊥,° Yan He,$ Xiaojing Huang,@ Hanfei Yan,@ Yong S. Chu,@ Simon J. L. Billinge,†,¶ and Ian K. Robinson*,†

† Condensed Matter Physics and Material Science Department, Brookhaven National Laboratory, Upton, NY 11973, USA

‡ Materials Science Division, Argonne National Laboratory, 9700 Cass Ave, Lemont, IL 60439, USA

¶ Department of Applied Physics and Applied Mathematics, Columbia University, New York, NY 10027, USA

§ Department of Physics, Columbia University, New York, NY 10027, USA

& Center for Functional Nanomaterials (CFN), Brookhaven National Laboratory, Upton, NY 11973, USA

// Key Laboratory of Materials Physics, Institute of Solid State Physics, Chinese Academy of Sciences, Hefei 230031, People's Republic of China

⊥ High Magnetic Laboratory, Chinese Academy of Sciences, Hefei 230031, People's Republic of China

° Collaborative Innovation Centre of Advanced Microstructures, Nanjing University, Nanjing 210093, People's Republic of China

$ Shanghai Synchrotron Radiation Facility (SSRF), Shanghai Institute of Applied Physics, Chinese Academy of Sciences, Shanghai, 201800, P. R. China

@ National Synchrotron Light Source-II, Brookhaven National Laboratory, Upton, NY 11973, USA

*Email: yue.cao@anl.gov; irobinson@bnl.gov

## ABSTRACT

Quasi-two-dimensional (quasi-2D) materials hold promise for future electronics because of their unique band structures that result in electronic and mechanical properties sensitive to crystal strains in all three dimensions. Quantifying crystal strain is a prerequisite to correlating it with the performance of the device, and calls for high resolution but spatially resolved rapid characterization methods. Here we show that using fly-scan nano X-ray diffraction we can accomplish a tensile strain sensitivity below 0.001% with a spatial resolution of better than 80 nm over a spatial extent of 100 $\mu$m on quasi 2D flakes of 1T-TaS$_2$. Coherent diffraction patterns were collected from a ~100 nm thick sheet of 1T-TaS$_2$ by scanning 12keV focused X-ray beam across and rotating the sample. We demonstrate that the strain distribution around micron and sub-micron sized 'bubbles' that are present in the sample may be reconstructed from these images. The experiments use state of the art synchrotron instrumentation, and will allow rapid and non-intrusive strain mapping of thin film samples and electronic devices based on quasi 2D materials.

## INTRODUCTION

The lattice degree of freedom profoundly affects the electronic properties of quasi-2D materials ranging from graphene to transition metal dichalcogenides (TMDCs). The stacking of layers,[1,2,3,4] in-plane strain[5] and strong electron-phonon coupling[6,7,8] amongst other factors determine the band structures, electron density of states as well as emergent charge orders [9,10,11] and superconductivity[12] in these materials. Moreover, quasi-2D materials are noted for their highly anisotropic mechanical properties, including superior strength[13,14] in the 2D plane, and have been used as electromechanical resonators.[15] Both observations put the measurement of strains in a critical position - such measurements will provide the basis for quantifying and simulating all other material properties, both electronic and mechanical.

To date, nano-scale strains in quasi-2D materials have been measured almost exclusively using nano-indentation,[13,14,16] where an AFM tip exerts a force normal to the 2D plane in contact mode before it punctures through. This is an intrusive method and requires free-standing quasi-2D materials. Moreover, deriving the Young's modulus and strain from these measurements often involves rigorous modelling and certain presumptions such as the morphology of the free-standing membrane.

To achieve truly model-independent, non-destructive and spatial-resolved measurements of the strain tensor, one need to turn to diffraction-based methods. The key perspective is to consider the entire quasi-2D crystal as consisting of a nanoscale array of 2D "tiles" each with their own lattice parameter and orientations. Strain in the continuous material can then be viewed as originating from seamlessly connecting neighboring tiles. The differences in the lattice parameters and relative orientations of each tile are normally referred to as 'mosaic' in the crystallography terminology, and give rise to the movements of the Bragg peak, as well as the geometrical differences in satisfying the Bragg condition. We could capture the variation of Bragg peak positions by scanning a small X-ray beam across the sample. As for the Bragg condition, it is possible to vary the sample orientation at each position in real space and identify the scattering angles that maximize the Bragg peak intensity. Up to date, such a capability has been demonstrated at the ID-01 (Microdiffraction imaging) beamline of the European Synchrotron Radiation Facility (ESRF) and elsewhere, but has been limited to fabricated toy models made of Si and related film structures.[17,18,19]

In this paper, we demonstrate strain mapping in quasi-2D 1T-TaS$_2$ flakes using X-ray nanodiffraction (nanoXRD). 1T-TaS$_2$ has been extensively studied as a model TMDC with multiple charge-ordered phases proximate to its onset of superconductivity. It is a Mott insulator with a $\sqrt{13} \times \sqrt{13} \times 3$ commensurate charge order below 200 K and undergoes a first-order metal-insulator transition into the nearly commensurate charge ordered phase. Another phase transition occurs at 350 K into the incommensurate charge ordered phase with the ordering vector (0.283, 0, 1/3).[12,20] Recent theoretical and experimental studies have suggested [6,7,11] that the inter- and intra-layer strains and electron-phonon coupling impact the macroscopic properties of this material, including the formation and stability of the charge orders. Here we focus on the strain mapping at room temperature, in the nearly commensurate charge ordered phase.

**EXPERIMENTAL METHODS**

Fig. 1 (a) depicts the X-ray nanodiffraction (nanoXRD) geometry. 1T-TaS$_2$ flakes were exfoliated mechanically and transferred to a 10 $\mu$m-thick Si substrate. All the data presented in this paper comes from a ~100 nm thick sheet of 1T-TaS$_2$. An Scanning Electron Microscopy (SEM) image of the nanosheet is also displayed in Fig. 1 (a). Details on material synthesis and sample preparation are described in the Supporting Information. The nanoXRD measurements were performed at room temperature and ~300 Torr helium gas (He) environment at the 3-ID (Hard X-ray Nanoprobe) beamline of the National Synchrotron Light Source-II, Brookhaven National Laboratory.[21, 22] Monochromatic X-ray with a photon energy of 12 keV and 80% coherence was focused by a Fresnel Zone Plate (FZP) onto the sample with a full-width-half-maximum (FWHM) of 80~100 nm. Tight X-ray focus with FWHM of ~ 50 nm was more often used at the beamline. A larger-than-usual focus was chosen in our case to avoid radiation damage to the charge order in 1T-TaS$_2$ (but not any of the structural Bragg peaks). The X-ray fluorescence from the sample was collected using a Vortex detector placed 90° relative to the incident X-rays. The diffracted X-rays, transmitted through the sample and the thin substrate, were collected using a pixel-array detector (Merlin with Medipix3 chip) in a forward scattering geometry. As the penetration depth of the X-ray is calculated to be ~ 3.5 $\mu$m, the diffracted X-ray photons come from the entire thickness of the sample rather than the top one or few atomic layers as in the surface-sensitive characterization methods. A detailed description of the beamline and the sample pre-alignment procedure is described in the Supporting Information and in ref. 22. A typical diffraction pattern is shown in Fig. 1(a). The diffraction intensity is distributed around a central dark region as is typical in coherent nano diffraction experiments.[23, 24, 25, 26] This central dark region originates from the use of a direct beam stop upstream of the FZP. The definition of diffraction angles in this paper follows the convention used in a six-circle X-ray diffraction geometry,[27] as illustrated in Fig. 1 (a). Two angles – sample tilt and azimuthal rotation (not drawn) – are limited by the instrumentation and kept constant throughout the experiment. The Bragg condition is uniquely defined by the sample rotation $\theta$ (with the rotation axis in the sample plane and normal to the experimental floor), the detector angles $\delta$ and $\gamma$.

## RESULTS AND DISCUSSION

In Fig. 1 (b-e) we show the maps of the Ta L-edge fluorescence intensity and the integrated peak intensities on the area detector at a number of Bragg reflections. Maps shown here and elsewhere in this paper are collected by rastering in the sample plane with a dwell time of 0.1 seconds using the 'fly-scan' mode.[28] The continuous scan mode is chosen to minimize the scanning overhead associated with stop-start of the piezo stages.[21] Most prominent in Fig. 1 (c-e) are the circular regions with a meridional bar, resembling 'coffee-beans'. The beans appear dark in their interior indicating a suppressed Bragg peak intensity. Integrated peak intensity maps at more Bragg and charge ordering peaks with different combinations of (H K L) have been obtained all showing the 'coffee beans' at the same location, apart from the different pointing of the 'meridional bar' which we will explain later in this paper. At any given spatial location on the nanosheet, the Bragg peaks could be refined using the same set of crystal orientation matrix defined locally on that location, and all with the lattice constant a = b = 3.35 Å, c = 5.86 Å, $\alpha = \beta = 90°$, $\gamma = 120°$. Since the Ta fluorescence does not show any changes in these regions, these observations collectively suggest that the missing intensities in the 'coffee-bean' regions are due to the sample locally not fully satisfying the Bragg condition as the rest of the bright regions.

To understand the nature of these 'coffee bean' profiles, we collected nanoXRD maps at a series of θ angles. The summed intensities from the (1, 0, 0) Bragg peak are displayed in Fig. 2 (a-h) through a rocking curve. As the θ angle increases monotonically, the bright parts develop from a single point (Fig. 2 (a)) to a deformed circle inside the left side of the coffee bean (Fig. 2 (b-c)). In Fig. 2 (d-e) the majority of the sample (which we refer to as the unperturbed region, as compared with those close to and within the 'coffee bean') meets the Bragg condition and exhibits maximum scattering intensity. From Fig. 2 (f) to (h), the right circumference of the coffee bean lights up, shrinks to a deformed circle and eventually a point inside the right side of the coffee bean. The θ values marked on top of Fig. 2 (a-h) and throughout the rest of the paper are θ angles relative to the Bragg angle of the unperturbed region. We further track the collected nanoXRD intensity vs. θ angle at a series of locations across one of the 'coffee beans' (Fig. 2 (i)). Each of these curves take on a Lorentzian line shape. The full width half maximum of each Lorentzian profile is around ~0.2°, corresponding to a correlation length of ~ 30 nm. The centroids of the peaks are seen to undergo a resonance-like oscillation, distinct at each spatial location.

In Fig. 2 for any given location in the projected 2D sample plane, the Bragg diffraction intensity reaches a maximum when the location satisfies a specific Bragg condition. We consider each location as a small tile with independent lattice parameters and orientations. To quantify these parameters at each location, we record the scattering geometry, including θ as well as the position of the diffraction peak on the detector, conventionally denoted by the angles δ (horizontal) and γ (vertical) (Fig. 1 (a)). In Fig. 3(a) we plot the θ angle with maximum Bragg intensity (denoted as $\theta_{max}$). At $\theta_{max}$ we further determine the center of mass of the diffraction pattern (as shown in Fig. 1(a)) and show the horizontal and vertical component of the center of mass in Fig. 3 (b) and (c) (denoted as $\delta(\theta_{max})$ and $\gamma(\theta_{max})$). Notably, the maps of $\delta(\theta_{max})$ and $\gamma(\theta_{max})$ are found to be essentially featureless at the position of the coffee-beans. The diffraction angle between the incident and scattered X-ray photons almost does not change across the two-dimensional sample plane. This diffraction angle is directly determined by the lattice parameter via the Bragg law. In our case, the lack of modulation in the diffraction angle of the (1, 0, 0) Bragg peak indicates all the local 1T-TaS$_2$ tiles share almost identical in-plane lattice parameters, a, even within the coffee-bean region. Also note the (0, 0, 1) axis of the 1T-TaS$_2$ tiles is known to be roughly normal to the supporting silicon wafer across the entire sample. Thus, any in-plane twist around the (0, 0, 1) axis could lead to the combination of $\delta(\theta_{max})$ and $\gamma(\theta_{max})$ to rotate along the Debye-Scherrer ring. The lack of such motion means the local 1T-TaS$_2$ tiles are almost azimuthally aligned. Hence the modulation in $\theta_{max}$ as shown in Fig. 3(a) could only come from minor relative tilts between adjacent 1T-TaS$_2$ tiles along the c axis. In this sense, our nanoXRD studies present an independent and self-contained survey of the local crystalline lattice.

We now piece together these tiles into a sheet of 1T-TaS$_2$. We expect the sample height function to be continuous, with the edges of neighboring tiles at the same sample height normal to the underlying silicon substrate. The difference in sample height between the mathematically infinitesimal tiles is

$$\frac{\partial h(x,y)}{\partial x} = tan\theta_{max}(x,y) \simeq \theta_{max}(x,y) \qquad (1)$$

where *h(x, y)* is the height of the sample along the z direction out of the quasi-2D plane and away from the underlying wafer. $\theta_{max}$ again is defined at each Bragg peak relative to the Bragg angle of the unperturbed region similar to elsewhere in this paper. The $x$ and $y$ directions are parallel to the detector $\delta$ and $\gamma$ angles at normal transmission respectively. We integrate in the quasi-2D plane

$$h(x,y) = \int^{x} \theta_{max}(x',y)dx' \qquad (2)$$

We derive the sample corrugation shown in Fig. 3(d). We can readily see that each of these coffee-bean shapes corresponds to a bubble-like, smooth one-sided deformation profile protruding above the substrate. Strictly speaking, equation (2) above only applies when the scattering plane is normal to the θ rotation axis (Bragg peak (1, 0, 0) for the specific sample). For all the Bragg peaks we explored, the scattering plane (defined by the incident X-ray and the combination of $\delta_{max}$ and $\gamma_{max}$ angles) is always normal to the direction of the fine 'meridional bar' connecting two halves of the coffee bean in the corresponding intensity map shown in Fig. 1. Where the scattering plane is tilted relative to the θ rotation axis, it could be mathematically shown that the 1D integration should be performed in the scattering plane.

Furthermore, these nanoXRD data are sufficient to allow direct calculation of the full strain tensor $\varepsilon_{ij}$ from the deformation profile. We plot all the nonzero elements of the strain tensor in Fig. 3 (e-i). Three of these terms, $\varepsilon_{xx}$, $\varepsilon_{yy}$, $\varepsilon_{xy}$ (and $\varepsilon_{yx}$), are proportional to $\theta_{max}^2$, or effectively zero if we only keep terms linear in $\theta_{max}$, reflecting extremely small changes in the lattice constants and d-spacings. This illustrates the high strain sensitivity of our nanoXRD method. $\varepsilon_{yz/zx}$ are linear in $\theta_{max}$ and are directly related to the sample rotation as we have so far ignored the thickness of our sample under the quasi-2D assumption.

We simulate the diffraction intensity of a Gaussian-like bubble protruding towards the incident X-ray at a series of θ rotations in Fig. 4 (a-e), with the scattering plane containing the x-axis. Fig. 4 (a-e) are consistent with our experimental observations. Fig. 4 (f-j) presents the diffraction intensity maps with scattering plane rotating 15° relative to the z-axis pointing out of the x-O-y plane. Note the 'meridional bar' connecting the two halves of the 'coffee-bean' is always orthogonal to, and rotates with the scattering plane. The nano tiles along the 'meridional bar' have their scattering vectors lying parallel to the scattering plane, and thus have maximum diffraction intensity. This effect has been experimentally observed in Fig. 1 (c-e).

The formation of these micron and sub-micron size bubbles could be attributed in part to the intrinsic 'softness' of these materials in the Mermin-Wagner sense. The Mermin-Wagner theorem[29] states that no long-range lattice order can exist in two dimensions, because it is unstable to thermal fluctuations. Accordingly, perfect 2D materials in principle should not lie completely flat in the 2D plane, but extend into the third dimension. As bulk materials become thinner towards the quasi-2D limit, it appears plausible that the materials also become more prone to the kind of corrugations presented in this work. Equally possible in our case, as the quasi-2D sheets are prepared at ambient pressure, a small volume of gas may have been trapped underneath the sheet. Our nanoXRD was performed at around 0.3 bar helium gas environment, possibly leading to a net pressure difference between the two sides of the sheet. Under this assumption, the height $h$ and the radii $r$ of a nano bubble in the 2D plane follows $R^2=r^2+(R-h)^2$ where $R$ is the curvature of the surface in 3D (Fig. 3(j) inset). In our case, the distribution of bubble sizes is consistent with a 3D surface curvature R ~ 0.3 mm (Fig. 3(j)). Given a pressure difference of 0.7 bar across the two sides of the bubble, Young's modulus can be estimated to be ~ 100 GPa for a strain of ~ 0.1%, which is three times the calculated value in bulk 1T-TaS$_2$,[30, 31] but less than 50% of the Young's modulus measured from ultrathin MoS$_2$.[16] Details of the calculation can be found in the Supporting Information.

For the small strains << 1% as measured in our sample, we observed that the distribution of the charge order is relatively uniform and follows the corrugation of the nanosheet. As the strain approaches a critical value around 3%, the electronic band structure deforms, and will eventually suppress the charge order as the electron density of states cross the Fermi level, leading to alternative charge scattering channels.[10]

**CONCLUSION**

We would like to compare our results with an electron diffraction version of strain mapping, which was recently demonstrated in refs. 32, 33. A major progress of that work, compared with previous scanning electron diffraction experiments, is to use a nano-sized electron source to avoid the complicated dynamical simulations needed for converging beam electron diffraction (CBED) studies. Still these electron diffraction methods are severely limited by the penetration depth of electrons and, furthermore, require samples to be in a free-standing geometry. In comparison, nanoXRD is well poised to study materials sealed as a device, with a top gate and/or a substrate. The spatial resolution of our nanoXRD approach is limited by the focused X-ray beam size (80 nm in this work), and could be improved using smaller X-ray beams (sub 10 nm in a state-of-the-art nanodiffraction facility such as the one where this work was performed[21, 22]), and also with the help of Bragg projection ptychographic reconstruction (1 nm as demonstrated in ref. 26). Indeed, our spatial resolution still lags behind electron and atomic force microscopes. On the other hand, as ~10 keV X-ray photons have much smaller momentum than those of electrons in a typical electron microscope, nanoXRD generally features better strain sensitivity than its electron counterpart. As quasi-2D materials get thinner, current electron diffraction experiments are inherently even more limited to quantify the sample morphology in the third dimension, preventing a full description of the strain tensor. Admittedly, approaches similar to ours have been demonstrated at ESRF, but only for pre-patterned model materials.[17, 18, 19] Here we have managed to experimentally extract strain and Young's modulus in realistic quasi-2D materials with superior spatial resolution even in the absence of ptychographic reconstruction, and will inspire future applications in obtaining crucial mechanical properties in these materials. The implications of our result will be significant when attempts are made in the future to assemble nano-scale heterostructures made from stacking quasi-2D materials on top of each other.[34] We suggest that nanoXRD could be a useful technique to understand the contacts formed between such quasi-2D materials and could have significant impact in that field.

**ASSOCIATED CONTENT**

Supporting Information is available online.

Bulk sample synthesis; nano flake preparation; X-ray nanodiffraction (nanoXRD); estimating Young's modulus

# AUTHOR INFORMATION

**Corresponding Author**

*E-mail: yue.cao@anl.gov; irobinson@bnl.gov

# ORCID

Yue Cao: 0000-0002-3989-158X

Tadesse Assefa: 0000-0003-3904-0846

Soham Banerjee: 0000-0001-9271-493X

Simon Billinge: 0000-0002-9734-4998

Ian Robinson: 0000-0003-4897-5221

**Author Contributions**

Y.C., I.K.R., S.B. and S.J.L.B. conceived the experiment. Y.C., T.A., S.B., Y.H., X.H., H.Y. and Y.S.C performed the nanoXRD measurement. Y.L., W.L. and Y.P.S. grew the bulk 1T-$TaS_2$ crystals. S.B., A.W., D.W. and A.P. further prepared the thin 1T-$TaS_2$ flakes used in the nanoXRD experiments. X.T. performed the atomic force microscopy characterization. Data analysis and interpretation was carried out by Y.C, I.K.R and S.J.L.B. Y.C. and I.K.R. wrote the majority of the paper with input from all coauthors.

**Notes**

The authors declare no competing financial interest.

# ACKNOWLEDGMENTS

Y.C. thanks Jing Tao, Jun Li and Chang-Yong Nam for valuable discussions, and acknowledges Stephan O. Hruszkewycz and Martin Holt for helpful earlier explorations. The work at Brookhaven National Laboratory and Columbia University was supported by the U.S. Department of Energy, Office of Basic Energy Sciences, Division of Materials Sciences and Engineering, under Contract No. DE-SC0012704. The work at Argonne National Laboratory was supported by the U.S. Department of Energy, Office of Basic Energy Sciences, under Contract No. DE-AC0206CH11357. This research used Beamline 3-ID HXN (Hard X-ray Nanoprobe) of the National Synchrotron Light Source II, a U.S. Department of Energy (DOE) Office of Science User Facilities operated for the DOE Office of Science by Brookhaven National Laboratory under Contract No. DE-SC0012704. Sample characterization was carried out using resources at the Center for Functional Nanomaterials, which is a U.S. DOE Office of Science Facility, at Brookhaven National Laboratory under Contract No. DE-SC0012704. S. B. acknowledges support from the National Defense Science and Engineering Graduate Fellowship (DOD-NDSEG) program. D. W. is funded by the Department of Energy under contract No. DE-SC0016703. Y.L., W.L. and Y.P.S. acknowledge support from the National Key Research and Development Program (Grant 2016YFA0300404), the National Nature Science Foundation of China (Grants 11674326, 11774352 and 11874357), the Joint Funds of the National Natural Science Foundation of China, and the Chinese Academy of Sciences' Large-scale Scientific Facility (Grant U1832141). Y. H. is supported by the Natural Science Foundation of Shanghai (17ZR1436800).

**FIGURES AND FIGURE CAPTIONS**

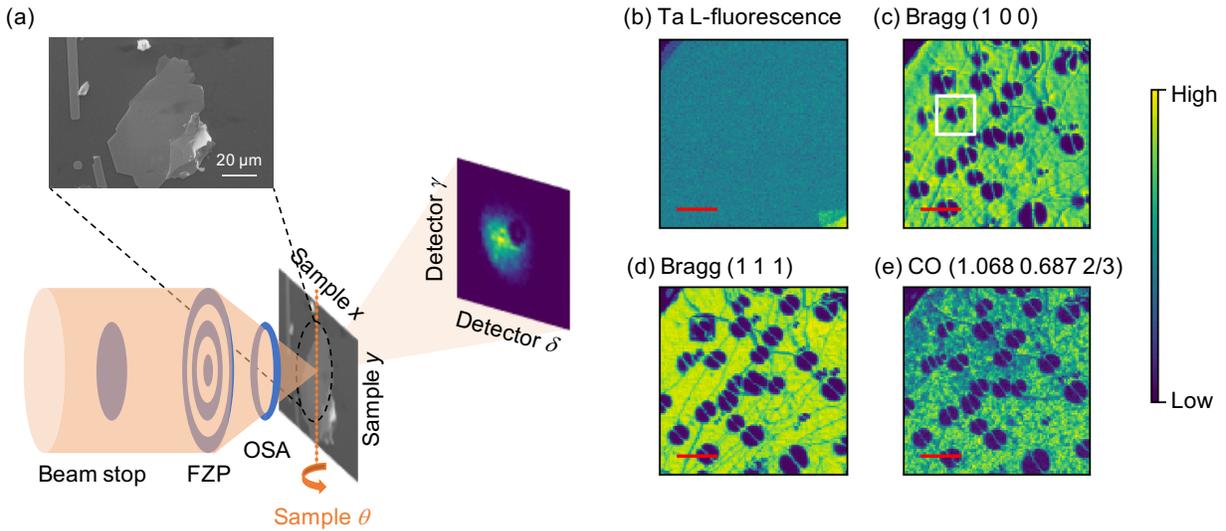

Figure 1: (a) The experimental setup. Monochromatic 12 keV X-rays with 80% coherence are focused using a Fresnel zone plate (FZP) and pass through an optical sorting aperture (OSA) onto the sample. The diffracted X-rays are collected using a pixel array detector behind the sample. The geometrical axes and coordinates used throughout the paper are as labeled. (b-e) shows the maps of (b) Ta L-edge fluorescence, (c-e) the total scattering intensity at Bragg (1, 0, 0), (1, 1, 1) and charge order (1.068, 0.687, 2/3) peaks. The red solid line corresponds to 5 $\mu$m. Each of (c-e) panels is collected at one respective $\theta$ angle, which is defined crystallographically as where most of the 1T-TaS$_2$ sheet satisfies the respective Bragg condition.

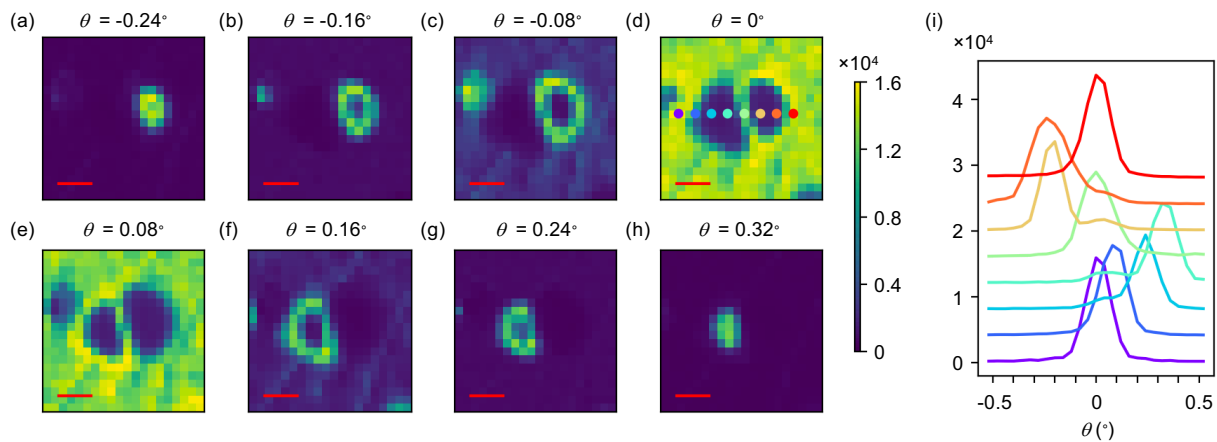

Figure 2: (a-h) NanoXRD diffraction intensity maps of the Bragg peak (1, 0, 0) at different $\theta$ angles measured from a 5 $\mu$m × 5 $\mu$m area as marked using a white solid box in Fig. 1. The red solid line corresponds to 1 $\mu$m. The $\theta$ values are defined relative to the Bragg angle of the 'unperturbed' or flat region (the bright part in panel (d) outside the coffee bean). (i) The summed diffraction intensity vs. $\theta$ angles at selected locations across the 'coffee bean'. The curves are offset vertically for clarity. The colors of curves correspond the locations marked in panel (d).

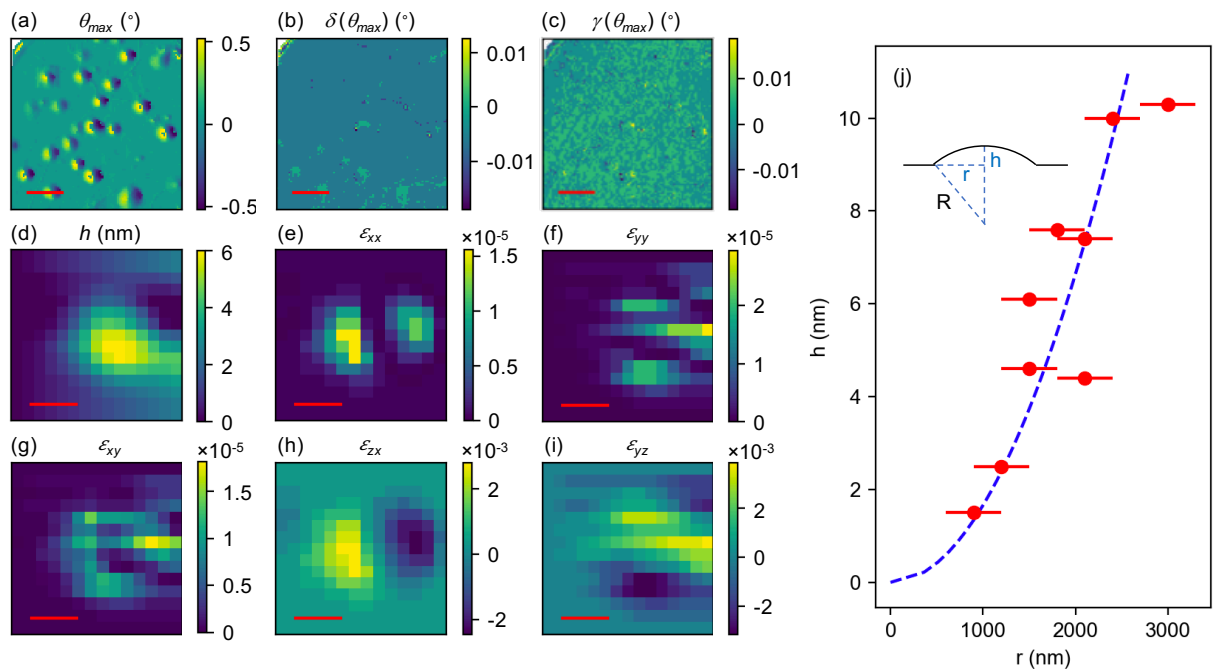

Figure 3: (a) The $\theta$ angle with maximum Bragg (1, 0, 0) diffraction intensity at different locations on the 1T-TaS$_2$ nanosheet, labeled as $\theta_{max}$. (b-c) The center of mass of the diffraction pattern at $\theta_{max}$ along detector angles $\delta$ and $\gamma$, labeled as $\delta(\theta_{max})$ and $\gamma(\theta_{max})$. All the units in (a-c) are in degrees, and the red scale bars correspond to 5 $\mu$m. (d) The height of the sample normal to the quasi-2D

plane, and normal to the underlying silicon wafer in the unit of nm. (e-i) Maps of all the nonzero and un-equivalent components of the elastic tensor calculated from the sample height in (d). Three of these terms, $\varepsilon_{xx}$, $\varepsilon_{yy}$, $\varepsilon_{xy}$ (e-g), are second order terms, while $\varepsilon_{zx}$ and $\varepsilon_{yz}$ (h-i), are linear terms. All the scale bars in (e-i) are 1 $\mu$m. (j) The radii $r$ and the height $h$ of a few nanobubbles. The inset shows a 1D schematic of a nanobubble. The error bar in the main panel is chosen to be the step size used in the scanning experiments. The dashed blue line is a fit for obtaining the surface curvature $R$ of the nanobubble.

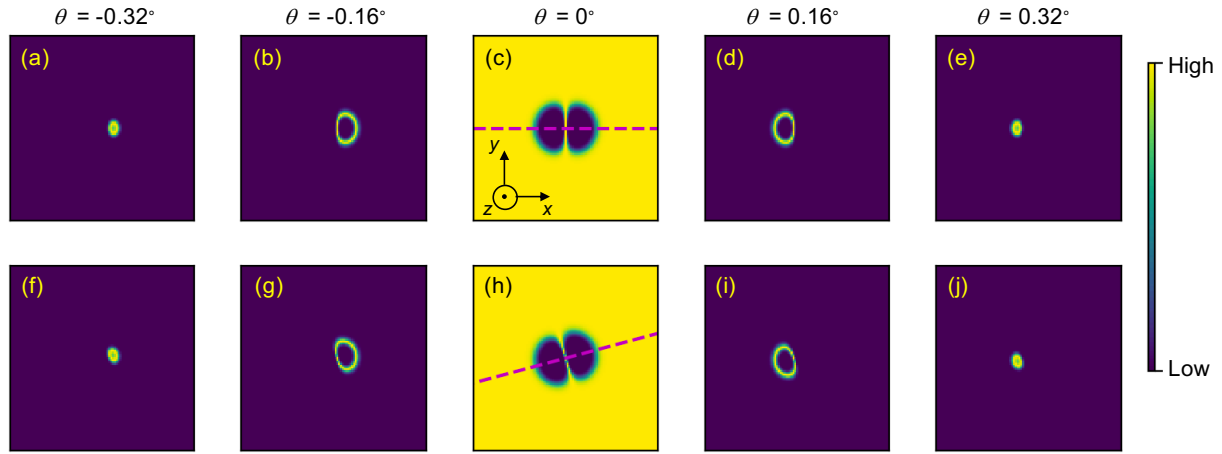

Figure 4: Simulated diffraction intensity maps from a 5 nm-high nanobubble at a series of $\theta$ angles. The scattering planes in (a-e) are in the $z$-$O$-$x$ and in (f-j) are 15° rotated counterclockwise around the $z$ axis. Projections of the scattering planes in the quasi-2D plane are marked using dashed purple lines in (c) and (h) respectively.

## Table Of Contents (TOC) graphic

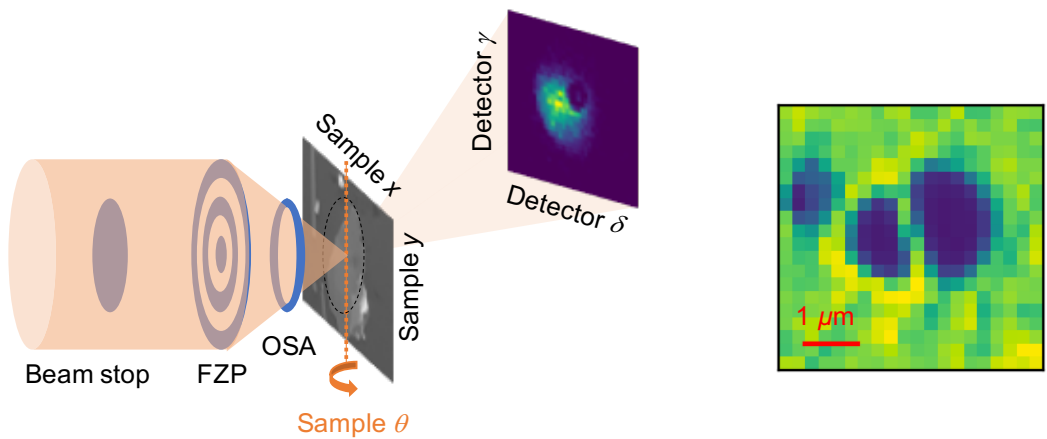

Supporting Information

# Complete Strain Mapping of Nanosheets of Tantalum Disulfide

Yue Cao,*,[†,‡] Tadesse Assefa,[†] Soham Banerjee,[†,¶] Andrew Wieteska,[§] Dennis Zi-Ren Wang,[¶] Abhay Pasupathy,[§] Xiao Tong,[&] Yu Liu,[//] Wenjian Lu,[//] Yu-Ping Sun,[//, ⊥, °] Yan He,[$] Xiaojing Huang,[@] Hanfei Yan,[@] Yong S. Chu,[@] Simon J. L. Billinge,[†,¶] and Ian K. Robinson*,[†]

[†] Condensed Matter Physics and Material Science Department, Brookhaven National Laboratory, Upton, NY 11973, USA

[‡] Materials Science Division, Argonne National Laboratory, 9700 Cass Ave, Lemont, IL 60439, USA

[¶] Department of Applied Physics and Applied Mathematics, Columbia University, New York, NY 10027, USA

[§] Department of Physics, Columbia University, New York, NY 10027, USA

[&] Center for Functional Nanomaterials (CFN), Brookhaven National Laboratory, Upton, NY 11973, USA

[//] Key Laboratory of Materials Physics, Institute of Solid State Physics, Chinese Academy of Sciences, Hefei 230031, People's Republic of China

[⊥] High Magnetic Laboratory, Chinese Academy of Sciences, Hefei 230031, People's Republic of China

[°] Collaborative Innovation Centre of Advanced Microstructures, Nanjing University, Nanjing 210093, People's Republic of China

[$] Shanghai Synchrotron Radiation Facility (SSRF), Shanghai Institute of Applied Physics, Chinese Academy of Sciences, Shanghai, 201800, P. R. China

[@] National Synchrotron Light Source-II, Brookhaven National Laboratory, Upton, NY 11973, USA

*Email: yue.cao@anl.gov; irobinson@bnl.gov

## Bulk sample synthesis

High-quality single crystals of 1T-TaS$_2$ were grown by the chemical vapor transport (CVT) method with iodine as a transport agent. The high-purity Ta (3.5 N) and S (3.5 N) were mixed in chemical stoichiometry and heated at 850°C for 4 days in an evacuated quartz tube. The harvested TaS$_2$ powders and iodine (density: 5 mg/cm$^3$) were then sealed in another quartz tube and heated for 2 weeks in a two-zone furnace, in which the source zone and growth zone were fixed at 900°C and 800°C, respectively. The tubes were rapidly quenched in cold water to ensure retaining of the 1T phase.

## Nano flake preparation

Thin 1T-TaS$_2$ flakes were micromechanically exfoliated onto the scotch tape. The 10 μm-thick silicon substrates were carefully inverted and placed atop large areas of the crystal. Subsequently, the substrate was slowly peeled off from its longer edge and examined under an optical microscope to identify suitable patches of crystal that had adhered to the substrate, taking into account the sample size, shape, and uniformity. We measured the thickness of one 1T-TaS$_2$ flake with the Park NX20-SPM Atomic Force Microscope using the non-contact mode, and calibrated it against the 80 nm-high Pt fiducial mark on the same silicon wafer. The thickness of all other flakes was determined using the Ta L-edge fluorescence intensity collected from the Vortex detector during the synchrotron experiment.

## X-ray nanodiffraction (nanoXRD)

Nanodiffraction measurements were performed using nano-Mii (Nanoscale Multimodal Imaging Instrument) at the Hard X-ray Nanoprobe Beamline of the NSLS-II.[1, 2] The 12 keV coherent incident X-rays are focused using a Fresnel zone plate with an outer ring width of 40 nm and outer diameter of 240 μm. This gives an incident X-ray convergence of 2.58 mrad, or 0.15°. The X-ray beam focused by the Fresnel zone plate (FZP) was prepared using a central beam stop and an order sorting aperture (OSA). Under the condition of this experiment, the focused beam size was about 80 nm. The photon flux is estimated around 2×10$^9$ photons/μm$^2$/sec at 12 keV. A pixel-array detector (Merlin, 55 μm/pixel, 512×512 pixels), positioned 0.5 m downstream of the sample, was used to collect the diffracted x-rays. A schematic drawing of the experimental setup as well as the diffraction geometry (including the $\theta$, $\delta$ and $\gamma$ angles) is displayed in Fig. 1(a) of the main text.

While the exfoliated 1T-TaS$_2$ often has edge facets indicating high-symmetry crystalline orientations, we need to self-consistently determine the crystal orientation during the nanoXRD experiments. This is achieved using a wide-angle Dexela CMOS X-ray detector (Dexela 1512NDT), mounted behind the Be exit-beam window of nano-Mii. The detector was pre-calibrated with Si wafer so that we could convert each pixel into the $\delta$ and $\gamma$ angles. By rotating (rocking) sample $\theta$ we captured both the crystal Bragg peaks and charge ordered superlattice peaks on the detector with the corresponding $\theta$, $\delta$ and $\gamma$ angles. We first filter the nanoXRD peaks by their two-theta values (i.e. d-spacings) and diffraction intensities, and then identify the crystal orientation by indexing the measured reflections using SPEC.[3] Nanodiffraction measurements with higher angular resolution and better detection sensitivity were carried out after removing the Dexela detector, allowing the diffracted x-rays to be transmitted to the Merlin detector. The nanoXRD data are collected using the fly-scan method as described in ref. 4. The data collection time per voxel is set to 0.1 seconds to prevent radiation damage. At least two consecutive fly-scans were taken at the same sample area to ensure there was no observable damage from the X-ray. Over the course of the 0.1 seconds, for a 300 nm step size, it takes 7.5µs for the piezopositioners to move ~300 nm. Ta L-edge fluorescence was collected using a Vortex detector positioned at 90° relative to the incident X-rays.

We collected simultaneously both the nanoXRD patterns (as those plotted on the area detector shown in Fig. 1 (a)) and the fluorescence X-rays from the sample while performing 2D imaging scans. X-ray fluorescence data were analyzed using PyXRF.[5]

**Estimating Young's modulus**
Following the definition in Fig. 3(j) inset of the main text, the surface stress $f$ per unit length at the circumference of the nanobubble of radii $r$ and height $h$ follows
$$f \sin\theta \cdot 2\pi r = \Delta p \cdot \pi r^2$$
where $R$ is the surface curvature, with $\sin\theta = r/R$. The strain is determined in Fig. 3 and is dominated by the off-diagonal term $\epsilon \sim 10^{-3}$. For a nano bubble with a thickness of $d$, Young's modulus is therefore
$$G = \frac{f}{\epsilon d} = \frac{\Delta p R}{2\epsilon d}$$
Here $\Delta p \sim 0.7$ bar, $R \sim 0.3$ mm, $d \sim 100$ nm. Thus $G \sim 100$ GPa.